\documentclass[reprint,showpacs,twocolumn,aps,prl]{revtex4-2}
\usepackage{amsmath,amsthm,amssymb}

\usepackage{graphicx}
\usepackage{subfig}
\usepackage{dcolumn}
\usepackage{bm}
\usepackage{caption}
\usepackage{balance}
\usepackage{bbm}
\usepackage{soul}

\usepackage{booktabs}
\usepackage{pgfplots}
\usepackage{graphicx}
\usepackage{lipsum,adjustbox}
\usepackage[flushleft]{threeparttable}
\usepackage{rotating}
\usepackage{booktabs}
\usepackage{multirow}

\usepackage[symbol]{footmisc}
\usepackage{stackengine}
\usepackage{tikz}

\usepackage{hyperref}
\usepackage{xurl}
\hypersetup{breaklinks=true}
\hypersetup{
    colorlinks=true,
    linkcolor=red,
    filecolor=red,
    urlcolor=red,
    citecolor=blue, 
}

\usetikzlibrary{calc, decorations.pathmorphing, decorations.pathreplacing, decorations.shapes}

\newcommand{\bra}[1]{\left\langle #1 \right|}
\newcommand{\ket}[1]{\left|#1\right\rangle}

\makeatletter
\newcommand*{\rom}[1]{\expandafter\@slowromancap\romannumeral #1@}
\makeatother

\begin{document}


\title{Unified trajectory criterion for quantum and classical non-Markovianity}

\author{\small Le Hu}
\email{le.hu@northwestern.edu}
\affiliation{Department of Physics and Astronomy, Northwestern University, Evanston, IL 60208, USA}
\author{\small and Archana Kamal}

\affiliation{Department of Physics and Astronomy, Northwestern University, Evanston, IL 60208, USA}
\date{\today}

\begin{abstract}
We propose a simple criterion for non-Markovianity: a quantum master equation is non-Markovian if and only if its \textit{trajectory set} contains a \textit{self-intersecting trajectory} (defined in the main text). Since self-intersection is invariant under time reversal, our criterion implies that Markovianity itself is time-reversal invariant: This property is not possessed by many existing criteria based on information flow or complete positivity. For a given quantum master equation, the set of trajectories generated from all initial states encodes its essential dynamical features. Thus, Markovianity can be determined directly from the trajectory set, reducing the problem to a general mathematical one: determing the non-Markovianity of the set itself, regardless of whether it originates from a quantum or classical process. Here, we solve this problem and classify the trajectory sets into three types: \textit{strictly Markovian}, \textit{initial-state Markovian}, and \textit{non-Markovian}. Through multiple examples, we compare our criterion with existing ones and show how they fail to capture Markovianity and non-Markovianity in various cases.
\end{abstract}
\maketitle
Quantum (non-)Markovianity has been a topic of intensive investigation in the past decade. Originating from the classical notion of Markov chains, a Markovian quantum process is characterized by ``memoryless-ness'', meaning that the future evolution of a state depends only on the current state, not on past states or evolutionary history. In contrast, a non-Markovian quantum process is more general and complex, where the evolution exhibits ``memory" effects. It is often associated with ``information backflow \cite{PhysRevLett.103.210401}''  and ``recoherence'', thus typically exhibiting distinct features from Markovian processes and potentially causing unexpected and undesirable results in quantum information processing. Consequently, numerous criteria have been proposed to determine what constitutes a Markovian or non-Markovian quantum process and to determine how to measure the degree of non-Markovianity. These include measures based on trace distance \cite{PhysRevLett.103.210401}, decay rates \cite{PhysRevA.89.042120}, Bloch volume \cite{PhysRevA.88.020102}, the quantum Fisher information \cite{PhysRevA.82.042103}, the divisibility of dynamics \cite{PhysRevLett.105.050403, PhysRevA.83.062115, PhysRevLett.112.120404}, the operational approach \cite{PhysRevLett.120.040405}, among others. See Refs.\,\cite{Rivas_2014, RevModPhys.88.021002, LI20181, Li_2019} for a review. However, different criteria often disagree on the extent of non-Markovianity, and sometimes even on whether a process is non-Markovian at all \cite{PhysRevA.89.042120} (Table \ref{tab:table1}). For instance, recently proposed spectral measure of non-Markovianity does not focus on information backflow and gives distinct answers from time-domain-based measures especially in the presence of correlated environments \cite{keefe2024quantifying}. Moreover, for many existing criteria, the implications of Markovianity versus non-Markovianity remain ambiguous, as such criteria are often tailored to specific features of quantum processes and thus may obscure—or even compromise—their consistency with the classical notion of the Markov property (see, e.g., Ref.~\cite{LI20181}).

In this Letter, we propose a new trajectory-based criterion for Markovianity and non-Markovianity. We show how the Markovianity of a quantum process can be solely determined from all possible trajectories it generates. As these trajectories are not limited to a quantum origin, the proposed criterion is universal, and hence applicable to quantum, classical, or even non-physical processes such as those encountered in biological systems or financial markets. This makes our proposed definition of Markovianity fully aligned with the classcial Markov property. We will motivate and detail our definitions with remarks, followed by multiple examples. Using these examples, we show how the existing criteria fail to capture the Markovianity and non-Markovianity in various cases.

\begin{figure}[!tbp]
\includegraphics[width=0.48\textwidth]{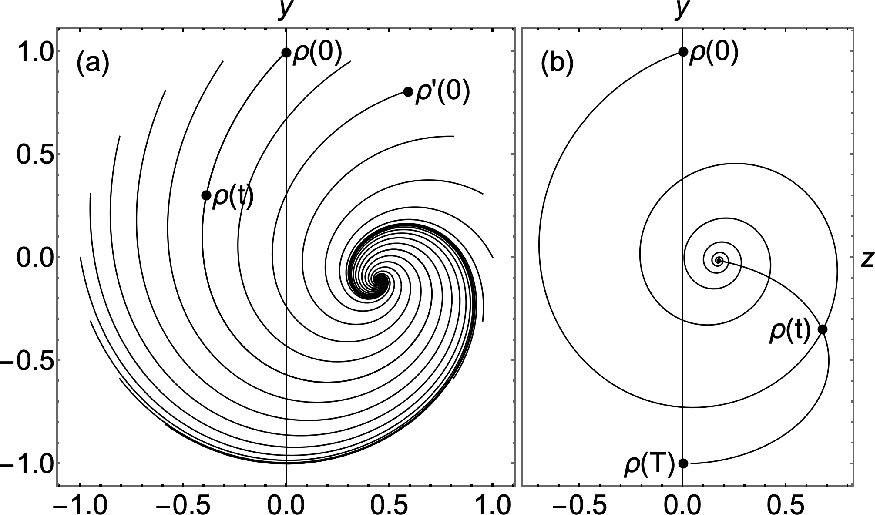}
\caption{
\label{fig:epsart1} Trajectories of Example 5 in the $y$–$z$ plane. Panel (a) Trajectories of different initial states with fixed $\gamma=1$, forming the trajectory set (the \textit{flow}). Panel (b): evolution trajectory with the initial state $\rho(0)=(x_0=0, y_0=0, z_0=1)$: $\gamma$ is held at $0.35$ until the state equilibrates, then slowly increased to $300$.  Without knowing $t$, one can always tell where $\rho(t)$ is evolving to in panel (a) but not in panel (b).} 
\end{figure}

\textbf{Motivation.}---Given an arbitrary quantum master equation (QME), the evolution history from an initial state $\rho(0)$ to some later state $\rho(t)$ forms a quantum trajectory. The trajectory, experimentally obtainable by quantum tomography \cite{murch2013observing, weber2014mapping, jullien2014quantum}, can be recorded in the Bloch vector form $\vec{x}$ in any finite $n$-dimensional space \cite{Bertlmann_2008}. Here, $\vec{x}$ is derived from $\rho= \frac{1}{n}(\mathbbm{1}+ \sum_{i=1}^{n^2-1} x_i \sigma_i)$ where $\sigma_i$ is the SU($n$) basis. For instance, for a two-level system $\rho(t)=\frac{1}{2}(\mathbbm{1}+x(t)\sigma_x+y(t)\sigma_y+z(t)\sigma_z)$, it can be recorded as $\rho(t):=(x(t),y(t),z(t))$. Now consider the cases where one exhausts all possible initial state $\{\rho(0)\}$. The evolution history generated from this family of initial states forms a \textit{trajectory set}, $\{\rho(t)\}=\{(x(t),y(t),z(t),\dots)\}$ [see, for example, Fig.~\ref{fig:epsart1}(a)]. Since the trajectory set contains all possible trajectories generated by the QME, by studying the behavior of the trajectory set $\{(x(t),y(t),z(t),\dots)\}$, we can determine the essential properties of the QME including whether it is Markovian or non-Markovian without needing explicit knowledge of the QME itself. This translates a quantum problem to a general mathematical problem, i.e., how to determine the Markovianity of a trajectory set $\{(x(t),y(t),z(t),\dots)\}$. In other words, we can say that the translated problem does not need to be related to quantum physics at all.

Before proceeding, it will be useful for us to briefly review the classical Markov chain. Consider a sequence of random variables $X_1, X_2, \dots$. It is called a discrete-time Markov chain if the sequence has the Markov property, meaning the probability of moving to the next state depends only on the current state, not on the previous states,
\begin{equation}
\begin{aligned}
	&\operatorname{P}\left(X_{n+1}=x \mid X_{n}=x_{n}, X_{n-1}=x_{n-1}, \ldots, X_{1}=x_{1}\right)\\=&\operatorname{P}\left(X_{n+1}=x \mid X_{n}=x_{n}\right).
	\end{aligned}
\end{equation}
If the transition probability $\operatorname{P}\left(X_{n+1}=x \mid X_{n}=x^\prime \right)=\operatorname{P}\left(X_{n}=x \mid X_{n-1}=x^\prime \right)$ is independent of time, then it is called a \textit{time-homogeneous} Markov chain, in a similar spirit to a time-independent master equation. Otherwise, it is called a \textit{time-inhomogenous} Markov chain, similar to time-dependent master equations. A Markov chain can also have finite memory, which is called a \textit{Markov chain of order m}, where $m$ is finite,
\begin{equation}
	\begin{aligned} & \operatorname{P}\left(X_{n}=x_{n} \mid X_{n-1}=x_{n-1}, \ldots, X_{1}=x_{1}\right) \\ = & \operatorname{P}\left(X_{n}=x_{n} \mid X_{n-1}=x_{n-1}, \ldots, X_{n-m}=x_{n-m}\right).\end{aligned}
\end{equation}
In other words, the future state depends on the past $m$ states. It is always possible to map a Markov chain of order $m$ to an ordinary Markov chain $Y_1, Y_2, \dots,$ by defining random variable $Y_n=(X_n, X_{n-1},\dots, X_{n-m+1})$ to include the past $m$ states of $X$. \textit{Hence, whether a process is Markov or not depends on how we define the ``state''.} 

Actually, it is well-known \cite{ross2014introduction} that the very same process, such as flipping a fair coin, can be Markovian when it is described by one set of random variables, e.g. $X_n=$ \{head, tail\} of the $n$th flip, but non-Markovian under another description, e.g. $Y_n=$ the number of times three consecutive tails have occurred. In other words, \textit{the Markov property is not an intrinsic property of the process itself but rather a property of how the process is described using a particular set of random variables.} Sometimes, it is totally fine to include past information in the ``current'' state. For instance, in classical physics, the state of a particle is characterized by its position $(x,y,z)$ and momentum $(p_x, p_y, p_z)$. The momentum contains past information as we need a short past history to measure momentum. \textit{When it comes to defining the Markovianity of a quantum process, we must be careful in defining what is a ``current state''.} 

In the following, we will provide our definition of Markovian processes, along with remarks and examples. We will restrict our attention to deterministic and continuous dynamics. For stochastic processes, we can always consider their ensemble average, which is again deterministic. Here, equations like the Schrödinger equation are considered deterministic, even though the measurement results may be random. Although the Markov property is initially applied in a stochastic context, we will apply it in a deterministic context here: deterministic evolutions can be regarded as special cases of stochastic evolutions.

\textbf{Definition 1.}---With full knowledge of the trajectory set, a quantum process is \textit{strictly Markovian} if, for any state $(x(t),y(t),z(t),\dots)$ along any trajectory at an \textit{unknown} time $t$, knowledge of the past trajectory provides no additional advantage in predicting the future:
$P(\rho(t+dt) | \rho(t)) =
	P(\rho(t+dt) | \rho(t), \rho(t-dt), \dots, \rho(0))$.
This is equivalent to saying with full knowledge on the trajectory set, we can determine $\dot{\rho}(t)$ solely from $\rho(t)$, without explicitly knowing $t$.

\textbf{Remark 1.}---It is critical to emphasize that time $t$ is \textit{not} included in the ``current state''. Namely, the current state is defined as $(x(t),y(t),z(t),\dots)$, not $(t, x(t),y(t),z(t),\dots)$. The reason is that if $t$ is explicitly known, then for any master equation $\dot{\rho}(t)=\mathcal{L}_t\rho(t)$ with known $\mathcal{L}_t$, the $\dot{\rho}(t)$ --- and hence future state $\rho(t+dt)$ --- is uniquely determined by $\rho(t)$. 
In other words, all quantum processes will become strictly Markovian and the whole problem becomes trivial. This should not be a surprise. Quantum evolutions before a measurement are deterministic, just like classical evolutions described by Newton’s laws. The future state at the next time-step depends only on the current state, not on the preceding states. 

\textbf{Definition 2.}---In the context of this paper, a trajectory $\rho(t)$ is called \textit{self-intersecting}, denoted as ``$\rho(t) \! \looparrowright$'', if there exists two different, finite times $t_1$ and $t_2$, such that $\rho(t_1)=\rho(t_2)$ and $\dot{\rho}(t_1)\neq\dot{\rho}(t_2)$.

\textbf{Remark 2.}---A self-intersecting trajectory may form a cross ``$\times$", a loop ``\scalebox{0.75}{$\bigcirc$}'', or go back and forth. If it forms a loop, note that the case where $\rho(t_1)=\rho(t_2)$ and $\dot{\rho}(t_1)=\dot{\rho}(t_2)$ does not count as self-intersection for the purposes of our discussion.

\textbf{Proposition 1.}---A Schr\"odinger equation or quantum master equation is strictly Markovian if and only if it is time-independent.
%

\textbf{Proof.}---Given a time-independent master equation $\dot{\rho}(t)=\mathcal{L} \rho(t)$, $\dot{\rho}(t)$ is always determined uniquely and solely from $\rho(t)$ by $\dot{\rho}(t)=\mathcal{L} \rho(t)$ regardless of $t$. Hence the master equation is strictly Markovian. Conversely, given a strictly Markovian master equation, since we can always determine $\dot{\rho}(t)$ solely from $\rho(t)$ without knowing $t$, the superoperator $\mathcal{L}$ must be time-independent. $\square$

\textbf{Remark 3.}---The definition of ``strictly Markovian'' coincides with the concept of the quantum Markov semigroup \cite{KOSSAKOWSKI1972247, fagnola1999quantum}. However, note that here and in what follows, we are \textit{not} concerned with whether the master equation is physical, i.e., whether the evolution from $\rho(0)$ to $\rho(t)$ can be described by complete positive and trace-preserving (CPTP) maps. This is because \textit{whether it is physical and whether it is Markovian are two separate considerations}. In other words, by the trajectory criteria, Markovianity or non-Markovianity is a universal property and should apply equally well to quantum, non-quantum, and even unphysical processes. See Example~1.

\textbf{Proposition 2.}---If a quantum process is strictly Markovian, then there is never an intersection between two trajectories at finite time, nor any self-intersecting trajectory. 

\textbf{Proof.}---By Proposition 1, a quantum process is strictly Markovian if and only if it is time-independent. For any time-independent first order linear ordinary differential equation, Picard-Lindel\"of theorem \cite{kent2012fundamentals} guarantees the uniqueness of the solution. Intersection of any kind will violate the uniqueness of the solution, because if one takes the intersection point as the initial state, then this implies multiple solutions. $\square$

\textbf{Remark 4.}---The converse is not true: no intersection of any kind alone does not necessarily imply a strictly Markovian process, since a time-dependent master equation can also exhibit this property; consider the counterexample, $\dot{\rho}=\gamma(t) (\sigma_z \rho \sigma_z -\rho)$ where $\gamma(t)=t$. It has solutions $x(t)=x_0 e^{-t^2/2},y(t)=y_0 e^{-t^2/2},z(t)=z_0$, which has no self-intersecting trajectory.

\textbf{Definition 3.}---With full knowledge of the trajectory set, a quantum process is called \textit{initial-state Markovian} if for any state $(x(0), y(0), z(0), \dots; x(t),y(t),z(t),\dots)$ drawn from any trajectory at an \textit{unknown} time $t$, knowledge of any past evolution history other than the initial state provides no additional advantage in predicting the future: $P(\rho(t+dt) | \rho(t), \rho(0)) =
	P(\rho(t+dt) | \rho(t), \rho(t-dt), \dots, \rho(0))$. This is equivalent to saying we can determine $\dot{\rho}(t)$ just from its initial state $(x(0), y(0), z(0), \dots)$ and current state $(x(t), y(t), z(t), \dots)$, without explicitly knowing $t$.

\textbf{Remark 5.}---The definition is motivated by the fact that one can always map a Markov chain of order $m$ to an ordinary Markov chain by including finite pieces of past information into the ``current'' state. In quantum information processing, we frequently do know about the initial state, hence it is reasonable to also include that information when we define the current state. 

\textbf{Proposition 3.}---A quantum process is initial-state Markovian if and only if there is no self-intersecting trajectory in its trajectory set.

\textbf{Proof.}---Let a quantum process described by the trajectory set $\{\rho(t)\}$ be initial-state Markovian. Suppose instead $\exists~\rho(t) \in \{\rho(t)\}$ such that $\rho(t) \! \looparrowright$. By Definition 2, this means $\exists~t_1, t_2 < \infty, t_1 \neq t_2$, such that $\rho(t_1)=\rho(t_2)$ and $\dot{\rho}(t_1)\neq\dot{\rho}(t_2)$. Since $\dot{\rho}(t_1)\neq\dot{\rho}(t_2)$, knowing both the initial state $\rho(0)$ and the current state $\rho(t_1)=\rho(t_2)$ cannot uniquely determine $\dot{\rho}(t)$. Hence $\{\rho(t)\}$ is not initial-state Markovian, contradicting the initial assumption.

%

Conversely, suppose $\forall \rho(t) \in \{\rho(t)\}$, $\rho(t)\! \not\looparrowright$. Knowing the initial state $\rho(0)$ will uniquely single out a trajectory $\rho(t)$ from $\{\rho(t)\}$. Since $\rho(t) \!\not\looparrowright$, $\nexists ~t_1, t_2 <\infty$ such that $ t_1\neq t_2, \rho(t_1)=\rho(t_2)$, and $\dot{\rho}(t_1)\neq\dot{\rho}(t_2)$. Hence, knowing the current state (say, $\rho(t_1)$, with $t_1$ unknown) will uniquely determine $\dot{\rho}(t)$ on this trajectory. Since knowing the initial and current state uniquely determines $\dot{\rho}(t)$, $\{\rho(t)\}$ is initial-state Markovian. $\square$

\begin{table*}[!tbp]
  \centering
  \caption{Comparison of the trajectory-based non-Markovianity criterion proposed in this work, with some other criteria in the literature. M: Markovian; SM: strictly Markovian; NM: non-Markovian; IM: initial-state Markovian.}
    \renewcommand{\arraystretch}{1.1}

    \begin{tabular}{ccccccc}
    \toprule
    Ex.&Highlight of Ex.& This paper     & Trace dist. \cite{PhysRevLett.103.210401} & Decay rates \cite{PhysRevA.89.042120} &CP-div. \cite{PhysRevLett.105.050403} & Bloch vol.\cite{PhysRevA.88.020102} \\
    \midrule
    1&$\dot{\rho}=-(\sigma_z \rho \sigma_z -\rho)$ &SM&NM*&NM*&NM* & NM*\\
    2&Certain unitary processes    &  NM     &   M  &  M &M&M\\
   3&Glued Markovian processes    &   NM    &     M  & M & M&M\\
    4&Negative rates but CPTP    &   IM    &  M     &NM  & NM&M\\
    5&Positive increasing rates     &  NM    &    M  &  M &M&M\\
    \bottomrule
    \end{tabular}%
        \label{tab:table1}%
        
    \begin{tablenotes}
      \small
      \item *Example 1 is unphysical; the NM results from applying the criteria naively.
    \end{tablenotes}

\end{table*}%

\textbf{Remark 6.}---If a quantum trajectory self-intersects, it means that the quantum state has returned to one of its past state, but via another route [i.e., $\dot{\rho}(t_1)\neq \dot{\rho}(t_2)]$. This change must be due to the change of the environment and/or its correlation with the system. That is, if the system and the environment as a whole is exactly the same as in the past, then the evolution of the system will also be exactly the same. Now since it is not exactly the same, it means something must have changed. Since the quantum state has not changed [i.e., $\rho(t_1)=\rho(t_2)$], then what changed must be the environment and/or its correlation with the quantum state. Such change is where the ``memory'' of the quantum state comes from. See Example 3.

The above interpretation is good if the Hamiltonian of the total system is time-independent. In systems with time-dependent total Hamiltonian, since the quantum state can be controlled to evolve along desired paths \cite{PhysRevA.110.062205, Alipour2020shortcutsto}, the source of quantum state's ``memory'' is the control Hamiltonian instead of the environment.

\textbf{Proposition 4.}---A quantum process is initial-state Markovian if the purity is always strictly decreasing, except for the stationary point where $\dot{\rho}=0$.

\textbf{Proof.}---Let $\rho= \frac{1}{n}(\mathbbm{1}+ \sum_{i=1}^{n^2-1} x_i \sigma_i)$, where $x_i$ is the component of Bloch vector and $\sigma_i$ is the SU($n$) basis. Since $\frac{d}{dt}\operatorname{Tr}\rho^2 = \frac{1}{n}\frac{d}{dt}|\vec{x}|^2<0$, there is no self-intersecting trajectory. By Proposition 3, the quantum process must be initial-state Markovian. $\square$

\textbf{Definition 4.}---A quantum process is \textit{non-Markovian} if it is not strictly Markovian nor initial-state Markovian.

\textbf{Proposition 5.}---A quantum process is non-Markovian if and only if it has a self-intersecting trajectory. (Proof follows from Proposition 2 and 3, and Definition 4.)


\textbf{Example 1.}---The QME $\dot{\rho}=-(\sigma_z \rho \sigma_z-\rho)$ is strictly Markovian given its time-independent nature, as per Proposition 1. Even though it leads to unphysical results, this fact alone should not prohibit us from discussing its Markovianity and the criteria should not suddenly fall apart. One can always restrict one's attention to only physical master equations if needed. 

\textbf{Example 2.}---Consider a time-dependent Schr\"odinger equation. By applying quantum control, we can always design the Hamtilonian $H(t)$ such that for some initial state $\ket{\psi(0)}$, the evolution trajectory $\ket{\psi(t)}$ self-intersects \cite{PhysRevResearch.5.033045, PhysRevLett.111.100502}. Hence even a unitary process can be non-Markovian under our definition. This is in contrast with all existing criteria on quantum Markovianity, which either regard all unitary processes as Markovian, or are not concerned with them at all. Note that for a closed quantum system, its Hamiltonian should generally be time-independent. Hence the time-dependency of the Hamiltonian implies that the system is actually in contact with some external source, such as a control field which maybe classical or quantum. In either case, it should not be surprising if the quantum state exhibits some memory.

The above example matters in the context of master equations with a time-dependent Hamiltonian. In such cases, it is not sufficient to just analyze the dissipative part of the QME, and one must also take into account the Hamiltonian to determine if the system dynamics are Markovian.

\begin{figure}[!tbp]
    \centering
    \includegraphics[width=0.29\textwidth]{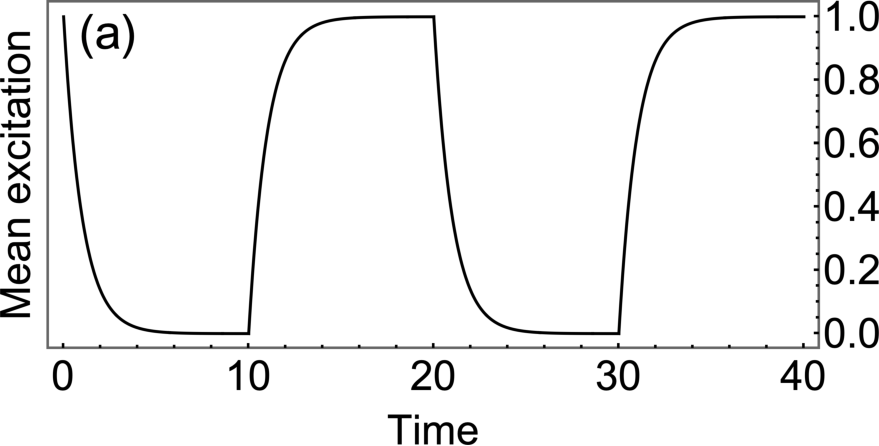}
    \hfill
    \includegraphics[width=0.14\textwidth]{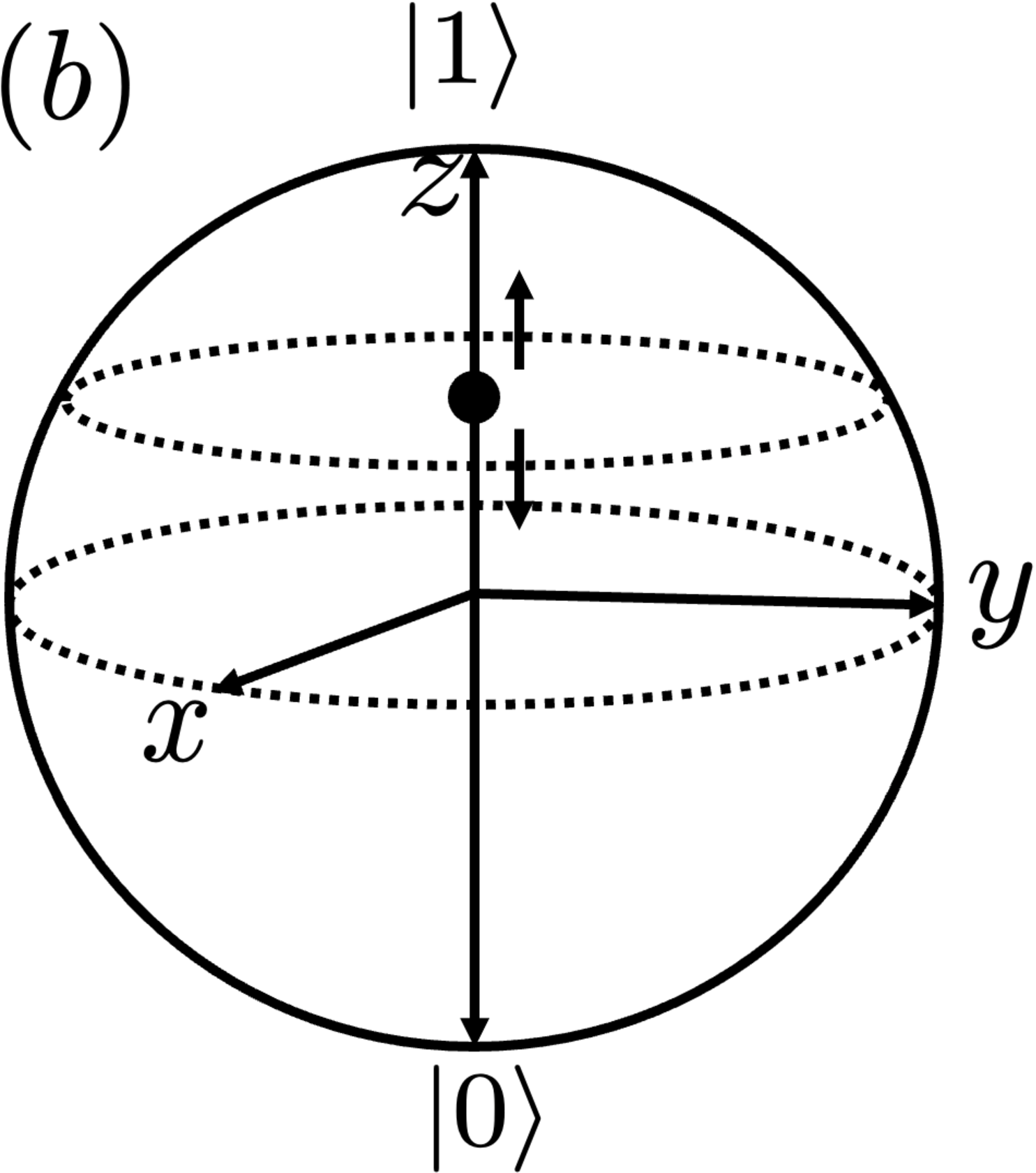}
\caption{\label{fig:epsart2}  (a) Mean excitation evolution for Example 3 with $T=10$. (b) Quantum state oscillation along the $z$-axis in Example 3. Initial state $\rho(0)=\ket{1}\bra{1}$.}
    \label{fig:combined}
\end{figure}

\textbf{Example 3.}---Consider a spin-1/2 particle flying through a spatially ordered sequence of baths, arranged as alternating cold and hot baths. The particle can exchange an excitation with each bath along its path. Suppose we are only interested in the spin space (rather than position space) of the particle, the dynamics of which can be described as
\begin{equation}
\begin{aligned}
	\dot{\rho}&=\gamma(t) (\sigma_- \rho \sigma_+ -\frac{1}{2}\{\sigma_+ \sigma_-,\rho\})\\
	&+[1-\gamma(t)] (\sigma_+ \rho \sigma_- -\frac{1}{2}\{\sigma_- \sigma_+,\rho\}),
\end{aligned}
\end{equation}
where $\sigma_-=\ket{0}\bra{1}$, $\sigma_+=\ket{1}\bra{0}$, and
\begin{equation}
\gamma(t)=\left\{\begin{array}{ll}1, & t \in[2 n T,(2 n+1) T] \text { where } n \in \mathbb{N} \\ 0, & \text { otherwise }\end{array}\right..
\end{equation}
$T\gg1$ is the constant time it takes to fly through each bath. Fig.\,\ref{fig:epsart2} shows the evolution of the mean excitation $\frac{\langle z\rangle+1}{2}$ and the trajectory with initial state $\rho(0)=\ket{1}\bra{1}$.

Most, if not all, existing criteria will regard the above master equation as Markovian, as evidenced by the fact that each ``piece" of the master equation is the canonical example of a Markovian process that every criterion on Markovianity must acknowledge. However, a quantum process glued from Markovian pieces may no longer be Markovian. 
To see this, consider the initial state $\rho(0)=\ket{1}\bra{1}$ and a given current state $\rho(t)=\frac{1}{2}(\ket{0}\bra{0}+\ket{1}\bra{1})$. Without knowing $t$ explicitly, one is unable to calculate $\dot{\rho}(t)$ because the appropriate value of $\gamma(t)$ to apply is unknown. However, if we know $\rho(t-\Delta t)=\frac{1+\epsilon}{2}\ket{0}\bra{0} +\frac{1-\epsilon}{2}\ket{1}\bra{1}$, where $\Delta t$ and $\epsilon$ are some small positive values, then we know we should take $\gamma(t)=0$. Since past information beyond the initial state improves future prediction, the process is non-Markovian. It is worth noting that if the state is instead defined as $\ket{\text{position space}} \otimes \ket{\text{spin space}}$, then the process becomes initial-state Markovian.


\textbf{Example 4.}---Consider $	\dot{\rho}=\frac{1}{2} \sum_{k=1}^{3} \gamma_{k}(t)\left[\sigma_{k} \rho \sigma_{k}-\rho\right]$,
with $\gamma_1(t)=\gamma_2(t)=1, \gamma_3(t)=-\tanh(t)$, and $\sigma_k$ being Pauli matrices. This example, first discussed in Ref.\,\cite{PhysRevA.89.042120}, 
is deemed non-Markovian by some measures and Markovian by others. The solution is given by $x_{j}(t)=\frac{1}{2}(1+e^{-2 t}) x_{j}(0)$,$ \quad x_{3}(t)=e^{-2 t} x_{3}(0)$ where $j=1,2$. From the solution it is evident that the trajectory set $\{(x_1(t), x_2(t), x_3(t))\}$ has no self-intersecting trajectory, hence the master equation is initial-state Markovian. This example, together with Example 1, indicates that a negative rate in the master equation does not necessarily imply non-Markovianity.

\textbf{Example 5.}---Consider the decay of a two level atom, $	\dot{\rho}(t)=-i\left[\sigma_{x}, \rho(t)\right]+\gamma(\sigma_{-} \rho(t) \sigma_{+}-\frac{1}{2}\left\{\sigma_{+} \sigma_{-}, \rho(t)\right\}),$
which differs from the regular case by replacing the Hamiltonian $\sigma_z$ with $\sigma_x$. The master equation is easily solvable and and admits a unique steady state, 	$(x_\infty =0, \quad y_\infty=\frac{4 \gamma}{\gamma^2+8}, \quad z_\infty=-\frac{\gamma^2}{\gamma^2+8}).$
Starting from an initial state such as $(x_0=0, y_0=0, z_0=1)$ and evolving under the influence of a small positive constant $\gamma$, the trajectory spirals into a steady state close to $(0,0,0)$. If $\gamma$ is then slowly increased, the quantum state remains near the instantaneous steady state, which traces out the right half of an ellipse. As $\gamma$ becomes large, the trajectory eventually completes the ellipse, producing a self-intersection [Fig.~\ref{fig:epsart1}(b)]. This shows that even a monotonically increasing positive $\gamma(t)$ can generate a non-Markovian process.

\textbf{Discussion.}---
Table \ref{tab:table1} summarizes the comparison of different criteria for non-Markovianity, with the third column being our classification. Our criterion represents both a conceptual and operational departure from existing time-domain signatures of non-Markovianity. Conceptually, it shares a key property with classical Markov chains: a classical Markov chain remains Markovian when reversed in time and on stable distribution \cite{norris1998markov}. Since trajectory intersections are invariant under time reversal, a process retains its classification as Markovian or non-Markovian under our criterion. By contrast, many widely used criteria—such as those based on information backflow, decay rates, or Bloch volume—flip sign or direction under time reversal and therefore fail to preserve Markovianity. Operationally, the trajectory criterion avoids the need for difficult optimization or piecewise integration, as required, for instance, by the trace-distance measure \cite{PhysRevLett.103.210401}. Although still tomographic in nature, our approach identifies non-Markovianity directly from the raw trajectories themselves.

We envision some possible extensions of the work presented here. One possibility is to quantify the degree of non-Markovianity based on trajectory-based criterion. A plausible approach is to define a measure of non-Markovianity as \textit{the ratio of self-intersecting trajectories in a given trajectory set}, with a high value (close to unity) indicating stronger non-Markovianity. This ratio denotes the probability of detecting the non-Markovianity of a (possibly unknown) system, assuming one prepares many different, random initial states and performs tomography of their evolution trajectories. For example, for a qubit interacting with a cavity in vacuum via a Jaynes-Cummings interaction, this ratio is unity, indicating the strongest non-Markovianity, while it is zero for Example 3, indicating the weakest non-Markovianity. It will also be interesting to explore quantum control protocols guided by this criteria which can restore Markovianity in time-dependent quantum systems.
\\\\
We thank Shengshi Pang for inspiring discussions. This work was supported by National Science Foundation under grant DMR-2508447.

\bibliography{apssamp5}

\end{document}